\documentstyle[prb,aps,multicol,psfig]{revtex}

\begin{document}
\draft                                           

\title{   Magnetic phases near the Van Hove singularity 
               in $s$- and $d$-band Hubbard model        }

\author{ Marcus Fleck, Andrzej M. Ole\'s,\cite{AMO} and Lars Hedin }

\address{ Max-Planck-Institut f\"{u}r Festk\"{o}rperforschung, 
          Heisenbergstrasse 1, D-70569 Stuttgart, Germany }

\date{March 21, 1997}
\maketitle

\begin{abstract}
We investigate the magnetic instabilities of the nondegenerate ($s$-band) 
and a degenerate ($d$-band) Hubbard model in two dimensions using many-body 
effects due to the particle-particle diagrams and Hund's rule local 
correlations. The density of states and the position of Van Hove 
singularity change depending on the value of next-nearest neighbor hopping 
$t'$. The Stoner parameter is strongly reduced in the $s$-band case, and 
ferromagnetism survives only if electron density is small, and the band is 
almost flat at small momenta due to next-nearest neighbor hopping. 
In contrast, for the $d$-band case the reduction of the Stoner parameter 
which follows from particle-particle correlations is much smaller and 
ferromagnetism survives to a large extent. Inclusion of local spin-spin 
correlations has a limited destabilizing effect on the magnetic states.
\end{abstract}


\begin{multicols}{2} 

\section{ Introduction }

Although the efforts to understand the microscopic origin of itinerant
magnetism have continued for over three decades, there is still no consensus
whether the nondegenerate ($s$-band) Hubbard model, originally introduced 
as a model for $d$-electrons in transition metals,\cite{Hub63} can serve 
as a simple model which describes itinerant ferromagnetism. On the 
mean-field level it seems to be a good starting point as the Stoner 
criterion predicts a ferromagnetic (F) ground state (GS) in a broad range 
of parameters. However, the inclusion of electron correlations destabilizes 
it in most situations.\cite{Hub63,Kan63,Ole82,vdL91,Han93} However, the
F state is stable in the limit of $U\to\infty$ up to a critical doping, 
being close to $\delta=0.29$ for a square lattice.\cite{vdL91,Han93} 
Recently, a few new mechanisms which stabilize ferromagnetism in the Hubbard 
model with moderate Coulomb repulsion $U$ have been proposed. They are 
realized either by extending the model by additional intersite Coulomb 
interactions,\cite{Hir89} or in the flat-band 
scenario.\cite{Lin87,Mie93,Tas95,Mul95,Pen96,Pie96} While the former 
mechanism might work in liquid hydrogen rather than in transition 
metals,\cite{Hir89} the question whether the spin-independent on-site 
Coulomb interaction $U$ alone can give ferromagnetism remains intriguing. 

Rather extreme situations in which the GS is F are encountered in low 
dimensional systems. First, high-spin GS are found in finite systems 
with open-shell electronic states, like in a tetrahedron,\cite{Vic84} 
and in some other few-atom clusters.\cite{Pas94} Second, the enhanced 
degeneracy at the Fermi level realized in one dimension by extended hopping 
stabilizes the F GS.\cite{Mie93,Tas95,Mul95,Pen96,Pie96} The mean-field 
analysis of Lin and Hirsch \cite{Lin87} suggests that the F instability is 
enhanced in two dimensions by the next-nearest neighbor hopping $t'$, 
and indeed the summation of the most divergent diagrams confirms the F 
instability at the Van Hove singularity (VHS) in two dimensions.\cite{Hlu96}
There are also indications from the enhanced stability of the Nagaoka state
by $t'$ at $U=\infty$ limit, that the kinetic energy changing slowly with 
electron filling favors ferromagnetism in a square lattice.\cite{Pre91}
 
Yet, in spite of this revived interest in the magnetic states of the 
nondegenerate Hubbard model, the magnetic states are realized in nature in 
degenerate $d$ bands of $3d$ transition metals. Therefore, we investigate 
here the F and antiferromagnetic (AF) instabilities of the $d$-band and
compare them with those found in the $s$-band for the same two-dimensional
(2D) lattice near the VHS. It is important to include the electron
correlations, if magnetic instabilities are considered. Using the 
Kadanoff-Baym technique of deriving conserving approximations, we find by
including particle-particle scattering a similar expression as postulated 
and used by Chen {\it et al.} in the $s$-band case.\cite{Che91} Note that
this approach is different from fluctuation exchange approximation 
(FLEX),\cite{Bic89,Fle89} and avoids the self-consistency in the conserving 
approximation, but nevertheless gives a magnetic structure 
factor of the same quality as the Monte-Carlo simulations in 
2D,\cite{Che91,Bul93} and in infinite dimensional Hubbard model.\cite{Fre95} 
In contrast, the experimental data suggest that the Hund's rule exchange 
interaction $J$ remains practically equal to its atomic value,\cite{vdM88} 
but the {\it atomic correlations} stabilize local moments for $J>0$. In the 
Hartree-Fock (HF) approximation these moments exist only in the symmetry 
broken phases, and they are absent in nonmagnetic states.\cite{Sto90} 
Thus, both particle-particle scattering and atomic correlations contribute 
to the reduction of the magnetic energy in the $d$-band, conventionally 
expressed by the Stoner parameter $I_d$. This motivated us to make a more 
extended study, in which we analyze the F and AF 
instabilities and discuss for the first time two questions: 
(i) Does ferromagnetism exist in the $s$-band model in a finite density 
range near the VHS? If it does, it might be possible to stabilize it also in 
three dimensions, provided a high density of states would exist at the Fermi 
level. 
(ii) How does the picture change when we go to the $d$-band case? 

The paper starts with the presentation of $s$-band and $d$-band Hubbard
models with next-nearest neighbor hopping in Sec. II. The $d$-band case is 
treated by a generalization of the treatment of the particle-particle 
scattering presented in Ref. \onlinecite{Che91}. We also include atomic 
spin-spin correlations by a local ansatz method and evaluate the 
renormalized Stoner parameters. The magnetic phase diagrams of the $s$-
and $d$-band are presented and discussed in Sec. III. The paper is concluded
in Sec. IV, where we also give estimations of the parameters used for
realistic transition metals.

\section{ The models and renormalized Stoner parameters } 

First, we consider a nondegenerate Hubbard ($s$-band) model on a square 
lattice with nearest ($t>0$) and next-nearest neighbor ($t'>0$) 
hopping,\cite{Lin87}
\begin{equation}
H_s = -t \sum_{\langle ij\rangle,\sigma} 
c_{i\sigma}^\dagger c_{j\sigma}\! + 
t'\sum_{\langle \langle ij \rangle\rangle,\sigma} 
c_{i\sigma}^\dagger c_{j\sigma} + 
U \sum_{i} n_{i\uparrow}\, n_{i\downarrow}.
\label{shubbard}
\end{equation}
An increasing $t'$ makes the kinetic energy, 
\begin{equation}
\varepsilon_{\bf k}=-2t(\cos{k_x}+\cos{k_y})+4t'\cos{k_x}\cos{k_y},
\label{dispersion}
\end{equation}
to increase slower from $\Gamma$ to $X(Y)$ point, and finally to become flat 
for $R=1$, where $R=2t'/t$. We do not consider the unrealistic cases with 
large $R>1$. Second, we study a degenerate $d$-band model, 
with a simplified intraorbital hopping,\cite{Ole83}
\begin{eqnarray}
H_d &=& -\,t \sum_{\langle i,j\rangle,\alpha,\sigma} 
c_{i\alpha, \sigma}^\dagger c_{j\alpha, \sigma} +\, 
t'\sum_{\langle\langle i,j \rangle\rangle,\alpha,\sigma} 
c_{i\alpha, \sigma}^\dagger c_{j\alpha, \sigma}  \nonumber\\
&+& U_0 \sum_{i, \alpha} n_{i\alpha, \uparrow}\, 
     n_{i\alpha, \downarrow} 
   +\left( U-{J\over 2}\right) \sum_{i,\alpha<\beta}  
     n_{i\alpha}\,n_{i\beta} \nonumber\\ 
&-& 2J\sum_{i, \alpha<\beta}{\bf S}_{i\alpha}\cdot {\bf S}_{i\beta}  
   + J\sum_{i,\alpha\not=\beta} 
c_{i\alpha,\uparrow  }^\dagger c_{i\alpha,\downarrow}^\dagger 
c_{i\beta ,\downarrow}^{}      c_{i\beta ,\uparrow  }^{}, 
\label{dhubbard}
\end{eqnarray}
where $n_{i\alpha}=\sum_{\sigma}n_{i\alpha,\sigma}$ and ${\bf S}_{i\alpha}=
\{S_{i\alpha}^x,S_{i\alpha}^y,S_{i\alpha}^z\}$ are density and spin operators 
for orbital $\alpha$ at site $i$, and $U_0=U+2J$. The $d$-bands have the same 
dispersion $\varepsilon_{\bf k}$ as in the $s$-band model (\ref{shubbard}). 

Following the approach of Kadanoff and Baym,\cite{Kad89} one may construct 
a 'conserving approximation' to the $d$-band Hubbard model which motivates
the approach by Chen {\it et al.}\cite{Che91} Therefore, we have 
considered a system coupled to an infinitesimal external field, 
${\bf b}_{i\alpha}=\left(b^x_{i\alpha},b^y_{i\alpha},b^z_{i\alpha}\right)$, 
\begin{equation}
H_d({\bf b})=H_d-\sum_{i\alpha}{\bf b}_{i\alpha}\cdot{\bf S}_{i\alpha},
\label{field}
\end{equation}
and derived a selfenergy by taking the functional derivative with respect 
to the full Green function $G$ of the Kadanoff and Baym potential 
$\Phi$, $\Sigma=\delta \Phi/\delta G$. 
In a finite field ${\bf b}_{i\alpha}$ one finds the Dyson equation for 
the one particle Green's function of the $d-$band Hamiltonian (\ref{field}),
\begin{equation}
{\hat G}\,^{-1}_{i j,\, \alpha}(\tau)=
{\hat G}\,^{0\; -1}_{ij,\alpha}(\tau) - \hat{\Sigma}_{ij,\alpha}(\tau) + 
{\bf b}_{i\,\alpha} \cdot {\hat {\bf \sigma}}\, \delta(\tau)\, \delta_{i,j}
\label{dyson}
\end{equation}
where ${\hat G}\,^{0}_{ij,\alpha}$ is the noninteracting (i.e., $U=J=0$)
(diagonal) Green function matrix, $\hat{\Sigma}_{ij,\alpha}$ is the 
(nondiagonal) selfenergy matrix labelled by spin indices, and 
$\hat{\bf \sigma}$ is a vector composed out of Pauli matrices. Due to the 
symmetry of the hypercubic lattice considered here, there are no 
interorbital hopping processes, and therefore the one-particle Green 
function ${\hat G}_{ij,\alpha}(\tau)$ is diagonal in orbital space.

The magnetic instabilities follow from the instabilities of the linear 
response function (the transverse susceptibility) to a local spin-flip 
excitation $(b^{-}_{i\alpha}=b^{x}_{i\alpha}-i\,b^{y}_{i\alpha})$, 
\begin{equation}
\chi\,^{\perp}_{i\,j,\, \alpha}(\tau-\tau') = 
\lim_{\epsilon\to 0^{+}}\sum_{\beta} \left. \frac{\partial\: 
G\,^{\uparrow\, \downarrow}_{j j,\, \alpha}
(\tau', \tau'+\epsilon)}{\partial\: b^{-}_{i,\, \beta}(\tau)} 
\right|_{{\bf b}=0} \; ,
\label{sus}
\end{equation}
which is determined by the full one-particle Green function $G$. 
Using a well-known identity we rewrite the functional derivative as,
\begin{eqnarray}
& &\left.\frac{\partial\: 
{\hat G}_{j j,\, \alpha}
(\tau',\tau'+\epsilon)}{\partial\: b^{-}_{i,\, \beta}(\tau)} 
\right|_{{\bf b}=0} = 
-  \int {\hat G}_{j i_{1},\, \alpha}
(\tau', \tau_{i_{1}})                             \nonumber \\ 
& &\hskip 1cm \times\left.\frac{\partial\: 
{\hat G}^{-1}_{i_{1} i_{2},\, \alpha}
(\tau_{i_{1}}, \tau_{i_{2}})}
{\partial\: b^{-}_{i,\, \beta}(\tau)} \, 
{\hat G}_{i_{2} j,\, \alpha}
(\tau_{i_{2}}, \tau'+\epsilon) \right|_{{\bf b}=0} \!\!\! ,
\label{iden}
\end{eqnarray}
where $\int \equiv \sum_{i_{\eta}}\, \int d\tau_{i_{\eta}}$ stands for the 
integration and summation over all internal variables $\{i_{\eta}\}$. 
Using the Dyson equation (\ref{dyson}), we find a formal integral equation 
for the transverse susceptibility, 
\begin{eqnarray}
&\chi&^{\perp}_{ij,\, \alpha}(\tau-\tau') = 
\chi^{\perp\, 0}_{i j,\, \alpha}(\tau-\tau')      \nonumber \\ 
&+& \sum_{\beta}\! \int \!{\rm Tr} \left[
{\hat \sigma}^{-}\, {\hat G}_{j i_{1},\, \alpha} (\tau', \tau_{i_{1}})  
\Gamma_{\alpha \beta}
(i_{1} \tau_{i_1},i_2\tau_{i_2}|i_3\tau_{i_3})\right. \nonumber \\  
& &\hskip 1cm \times\left.\left. {\hat G}_{i_2 j,\alpha}(\tau_{i_2},\tau')
\right]\chi^{\perp}_{i\,i_{3},\beta}(\tau-\tau_{i_3})
\right|_{{\bf b}=0} \!\!\! .
\label{trsus}
\end{eqnarray}
where ${\hat\sigma}^-={\hat\sigma}^x-i{\hat\sigma}^y$ is a Pauli matrix,
and the {\it effective two particle} interaction is defined as
\begin{equation}
\Gamma_{\alpha \beta}(i_1\tau_{i_1}, i_2\tau_{i_2}| i_3\tau_{i_3})
\equiv \lim_{\epsilon\to 0^+}\frac{\partial{\hat \Sigma}_{i_1 i_2,\,\alpha}
(\tau_{i_{1}}, \tau_{i_{2}})}
{\partial G^{\uparrow \downarrow}_{i_3 i_3,\, \beta}
(\tau_{i_3}, \tau_{i_3}+\epsilon)}\; .
\label{effint}
\end{equation}
One observes immediately, that an RPA-like expression for the transverse 
susceptibility would be obtained, if the selfenergy, 
$\hat{\Sigma}_{ij,\alpha}(\tau)$, 
were local in space and time, as is, e.g., the HF-selfenergy.  

\begin{figure}
\centerline{\psfig{figure=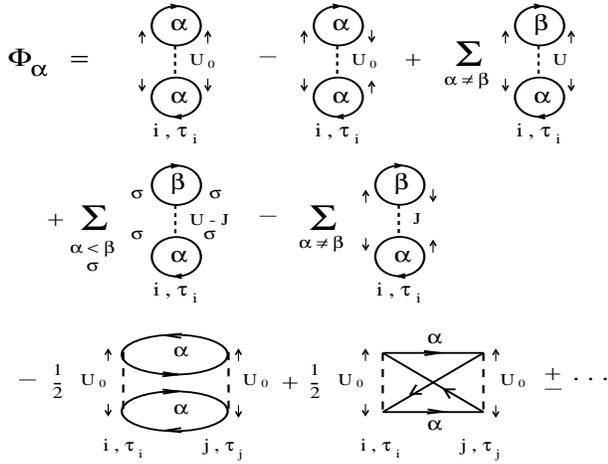,height=2.4in,width=3.2in}}
\narrowtext
\smallskip
\caption
{Schematic representation of the diagrammatic expansion of the Kadanoff-Baym 
potential part which originates from electron-electron interaction with a 
representative $d$-orbital, $\Phi_{\alpha}$. The first and second order
diagrams are shown; higher order diagrams include multiple particle-particle
scattering and dress the intraorbital Coulomb interaction $U_0$ further.}
\label{diagrams}
\end{figure}
We will now use the Kadanoff-Baym technique which may be used to obtain
conserving approximations.\cite{Kad89} First, we rewrite the expression 
for $H_d$ (\ref{dhubbard}) in a more explicit form which separates the 
spin-spin interaction into transverse and longitudinal terms,
\begin{eqnarray}
H_d &=& -\,t \sum_{\langle i,j\rangle,\alpha,\sigma} 
c_{i\alpha, \sigma}^\dagger c_{j\alpha, \sigma} +\, 
t'\sum_{\langle\langle i,j \rangle\rangle,\alpha,\sigma} 
c_{i\alpha, \sigma}^\dagger c_{j\alpha, \sigma}  \nonumber\\
&+& U_0 \sum_{i\alpha} n_{i\alpha\uparrow}n_{i\alpha\downarrow} 
 - J\sum_{i,\alpha<\beta,\sigma}
   c^\dagger_{i\alpha, \sigma}c^{}_{i\alpha,-\sigma}
   c^\dagger_{i \beta,-\sigma}c^{}_{i \beta, \sigma}      \nonumber\\
&+&\sum_{i,\alpha<\beta,\sigma}  
   \left[      U n_{i\alpha,\sigma}n_{i\beta,-\sigma}
        + ( U-J )n_{i\alpha,\sigma}n_{i\beta, \sigma} \right] \nonumber\\ 
&+& J\sum_{i,\alpha\not=\beta} 
    c^\dagger_{i\alpha,  \uparrow} c^\dagger_{i\alpha,\downarrow} 
    c^{     }_{i \beta,\downarrow} c^{     }_{i \beta,  \uparrow}. 
\label{otherdhubbard}
\end{eqnarray}
The diagrams to lowest order in the Kadanoff-Baym functional $\Phi$ (see
Fig. \ref{diagrams}) are obtained by taking GS expectation values of all 
possible contractions of the operators in $H_d$. Since we have a spin-flip 
term due to the infinitesimal field ${\bf b}_{i\alpha}$, also spin-flip 
Green functions appear. In choosing higher order terms, we have only kept 
the diagrams corresponding to particle-particle scattering, similarly as 
in the FLEX method in Ref. \onlinecite{Fle89}. 

For Kadanoff-Baym diagrams we only keep those which contribute both to the 
Green's function and to the susceptibility. In this way we avoid 
inconsistencies when we calculate phase boundaries from total energy, using 
the Green's function, and from finding singularities in the susceptibility. 
This means that we only keep diagrams of second order in 
$\langle c^\dagger_{i\alpha, \sigma} c^{}_{i \beta,\sigma'}\rangle$, 
since diagrams of higher and lower order give vanishing contribution when we 
take the two functional derivatives and the limit of zero external field,
${\bf b}_{i\alpha}\rightarrow 0$. The leading diagrams of a consistent 
theory are shown in Fig. \ref{diagrams}. The generating functional for the 
selfenergy is obtained by summing up to infinite order the class of diagrams 
in the particle-particle channel, which leads to an alternating geometric 
series, shown in Fig. \ref{diagrams}. One then integrates over all internal 
variables and sums over orbitals, $\Phi=\sum_{\alpha}\,\int\Phi_{\alpha}$. 

Short-range electron-electron correlation effects, neglected within the RPA, 
consist of local particle-particle scattering processes. Here we use a {\em 
local approximation}, and thus the contribution of each higher order diagram 
to $\Phi_{\alpha}$ is $\propto\delta_{i,j}$. 

\begin{figure}
\centerline{\psfig{figure=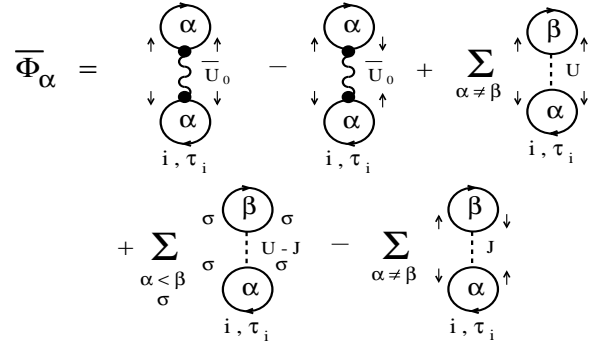,height=1.8in,width=3.0in}}
\narrowtext
\smallskip
\caption
{Schematic representation of the diagrammatic expansion of the effective 
Kadanoff-Baym potential, ${\bar \Phi}_{\alpha}$, with the effective 
intraorbital Coulomb interaction ${\bar U}_0$, defined by Eq. 
(\protect{\ref{ubar}}).}
\label{ediagrams}
\end{figure}
Compared to other dynamical electronic processes, particle-particle 
scattering can by considered as instantaneous process. Focusing on the 
local particle-particle kernel, 
$\chi^{pp}_{ii,\alpha}(\tau_{i}-\tau_{j})=
G^{  \uparrow  \uparrow}_{ii,\alpha}(\tau_{i}-\tau_{j})\, 
G^{\downarrow\downarrow}_{ii,\alpha}(\tau_{i}-\tau_{j})$, 
shown in the second-order diagrams of Fig. \ref{diagrams}, it is reasonable 
to write $\chi^{pp}_{ii,\alpha}(\tau_{i}-\tau_{j}) \approx 
          \chi^{pp}_{ii,\alpha}(\omega=0)\,\delta(\tau_{i}-\tau_{j})$, where 
\begin{equation}
\chi\,^{pp}_{ii,\alpha}(0)=\frac{1}{\beta} \sum_{n} 
G^{  \uparrow  \uparrow}_{ii,\alpha}(i\omega_{n})\, 
G^{\downarrow\downarrow}_{ii,\alpha}(-i\omega_{n})\; ,
\label{scver}
\end{equation}
with $\beta=1/k_BT$. Within this approximation one may replace the 
interaction between two electrons in $|\alpha\uparrow\rangle$ and 
$|\alpha\downarrow\rangle$ states, shown by higher order diagrams in Fig. 
\ref{diagrams}, by the effective interaction ${\bar U}_0$, and we can map 
$\Phi$ onto an effective Kadanoff-Baym potential, 
${\bar\Phi}=\sum_{\alpha}\,\int{\bar\Phi_{\alpha}}$, shown by its 
diagrammatic representation in Fig. \ref{ediagrams}. The interaction vertex 
in the effective Kadanoff-Baym potential obeys the following equation,
\begin{equation}
{\bar U}_0=U_0-U_0\,\chi^{pp}_{\alpha}(0)\,{\bar U}_0\; ,
\label{urn}
\end{equation}
where $\chi^{pp}_{\alpha}(0)=\chi^{pp}_{ii,\alpha}(0)$ for a translationally
invariant system, which is easily solved to give an effective intraorbital 
Coulomb interaction in $d$-band,
\begin{equation}
{\bar U}_0=\frac{U_0}{1+U_0\chi^{pp}_{\alpha}(0)}\;.
\label{ubar}
\end{equation}

It is now straightforward to calculate the selfenergy in the one-particle 
Green's function. Performing explicitly the functional derivatives, 
$\Sigma=\delta\Phi/\delta G$, we find the local selfenergy matrix,
\begin{equation}
\hat{\Sigma}_{ij,\alpha}(\tau-\tau') = \left(\begin{array}{cc} 
\Sigma\,^{  \uparrow   \uparrow}_{i\alpha} & 
\Sigma\,^{  \uparrow \downarrow}_{i\alpha} \\
\Sigma\,^{  \uparrow \downarrow}_{i\alpha}\,^* & 
\Sigma\,^{\downarrow \downarrow}_{i\alpha}\end{array} \right)
\delta(\tau-\tau')\,\delta_{i,j} \, ,
\label{self}
\end{equation}
with elements
\begin{eqnarray}
\Sigma_{i\alpha}^{\sigma\sigma}& = &
[{\bar U}_0\langle n_{i\alpha,-\sigma}\rangle+\sum_{\beta\not=\alpha}
\left( U\langle n_{i\beta}\rangle 
       -J\langle n_{i\beta,-\sigma}\rangle\right)] , \nonumber \\
\Sigma_{i\alpha}^{\uparrow \downarrow}& = &-
[{\bar U}_0\langle S^{-}_{i\alpha}\rangle - 
J \sum_{\beta\not=\alpha}\langle S^{-}_{i\beta}\rangle].
\label{sigma}
\end{eqnarray}
The adopted approximation to the selfenergy has the same functional form 
as the usual Hartree-Fock (HF) approximation, and therefore we call it 
generalized HF (GHF) approximation. Self-consistency for the presented 
$\Phi$-derivable GHF approximation is required only for the GHF occupation 
numbers $\langle n_{i\alpha,\sigma}\rangle$.

Next, an RPA-like expression for the transverse susceptibility is obtained, 
using the Green function with the selfenergy (\ref{sigma}) in Eqs. 
(\ref{effint}) and (\ref{trsus}),
\begin{equation}
\chi\,^{\perp}_{\alpha}({\bf q},0)= {\chi^0_{\alpha}({\bf q},0)\over
1 -I_d \chi^0_{\alpha}({\bf q},0)}, 
\label{chi}
\end{equation}
which we will call generalized (GRPA), since it has the same functional form 
as in RPA. In the $s$-band case the above formula contains $I_s={\bar U}$ 
(instead of $I_d$), where ${\bar U}$ is defined by the same renormalization 
due to the particle-particle vertex as in Eq. (\ref{urn}), but with $U_0$ 
replaced by $U$, and gives an excellent agreement with the Monte-Carlo 
data.\cite{Che91} There seems no reason that this approximation should not 
work well also in the $d$-band case. 

As an important difference to RPA, the HF value of the Stoner parameter 
for $d$-band model, $I_{\rm HF}=(U+6J)/5$, is now replaced by the 
renormalized Stoner parameter, 
\begin{equation}
I_d=\frac{1}{5}({\bar U}_0+4J).
\label{stoner}
\end{equation}
We note that the Stoner parameter in the $s$-band is just equivalent to the
renormalized value of $U$, $I_s={\bar U}$. 

\begin{figure}
\centerline{\psfig{figure=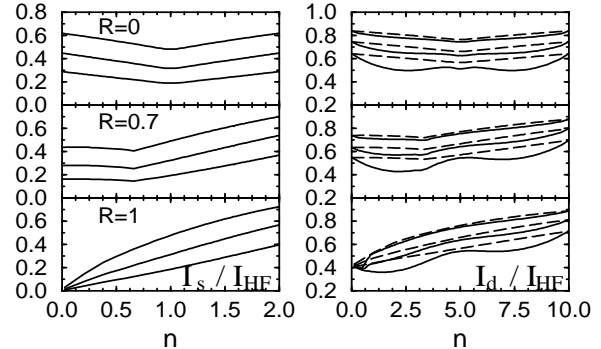,height=1.8in,width=3.0in}}
\narrowtext
\smallskip
\caption
{Stoner parameters $I_s/I_{HF}$ for the $s$-band (left), and $I_d/I_{HF}$ 
for the $d$-band (right) as a function of filling for different $R$-values. 
The three full-drawn curves in each panel refer to $(U+6J)/W=0.5$, $1.0$ 
and $2.0$ from top to bottom. The dashed curves in the $d$-band case are 
results with spin-spin correlations (from local ansatz) included.}
\label{reno}
\end{figure}

The reduction of the Stoner parameter due to the screening of the 
intraorbital Coulomb interaction $U_0$ may be substantial in a $d$-band, but
not quite as big as in the $s$-band case, as we show below. In contrast, 
the screening of the exchange interaction $J$ is 
provided by similar expressions which involve the interorbital transitions 
on the same site, $G^{\sigma\sigma}_{i\alpha,i\beta}$, and is thus of second
order in $\langle c^{\dagger}_{i\alpha,\sigma}c^{}_{i\beta,\sigma}\rangle$. 
If the interorbital hopping vanishes (as it does for hypercubic lattices), 
$J$ is unscreened; otherwise this screening is expected to be small. This 
is confirmed by the values of $J$ deduced from the experimental data in 
transition metals which are close to the atomic values.\cite{vdM88} We note 
that the renormalization of $I_d$ is therefore substantially weaker 
than in the $s$-band Hubbard model (\ref{shubbard}) (Fig. \ref{reno}), where 
$I_s={\bar U}$, and $\bar U$ is obtained from Eq. (\ref{urn}) with $J=0$. 
In the 2D Hubbard model one finds that $I_s$ is only weakly dependent on 
the band filling $n$, with a minimum at the filling which corresponds to 
the VHS, but is {\it finite} as long as $R<1$. With increasing $R$, 
the VHS moves towards the lower band edge, the reduction of $I_s$ 
gets stronger, and one finds that $I_s\to 0$ for $R\to 1$. Such a strong 
renormalization of $I_s$ follows from the singular behavior of the 
particle-particle vertex. 

The Stoner parameter is further decreased in the $d$-band case by the atomic 
interorbital correlations.\cite{Sto90} This results from the formation of 
local moments which can be built into the GS by modifying a 
single-Slater-determinant HF wave function $|\Psi_{\rm HF}\rangle$ 
into,\cite{Sto96}
\begin{equation}
|\Psi_0(\{n_{i\alpha,\sigma}\})\rangle=
\exp( -\sum_m\eta_mO_m )|\Psi_{\rm HF}(\{n_{i\alpha,\sigma}\})\rangle,
\label{ansatz}
\end{equation}
where $\eta_m$ are variational parameters, and $O_m$ are local operators
which in the case of a $d$-band have the form,
\begin{eqnarray}
O_{i,\alpha\beta}^{(n)}&=&n_{i\alpha} n_{i\beta}, \nonumber \\
O_{i,\alpha\beta}^{(s)}&=&{\bf S}_{i\alpha}\cdot {\bf S}_{i\beta},
\label{localo}
\end{eqnarray}
and describe local density ($O_{i,\alpha\beta}^{(n)}$) and spin 
($O_{i,\alpha\beta}^{(s)}$) correlations, respectively. The contribution of 
spin correlations to the magnetic energy follows from a comparison of the 
energy obtained with the wave function 
$|\Psi_0(\{n_{i\alpha,\sigma}\})\rangle$ given by Eq. (\ref{ansatz}) with
that found with density correlations only.\cite{Sto90} We note 
that the HF and correlated wave functions in Eq. (\ref{ansatz}) are
obtained for the same electronic distribution. The Stoner parameter $I_d$ 
is obtained as a derivative of the interaction energy with respect to
magnetization squared, and is additionally reduced by up to $12\%$ of 
$I_d$, if $(U+6J)/W=1$ (Fig. \ref{reno}). Therefore, one finds at 
$(U+6J)/W=1$ that $I_d/I_{\rm HF}\simeq 0.65$ which agrees well with the 
values deduced within a realistic model for $3d$ transition 
metals.\cite{noteid} Interestingly, at $R=0$ the largest corrections due 
to spin-spin correlations are found at $|n-5|\simeq 2.5$, which indicates 
that the difference between local moments in the paramagnetic (P) and 
weakly F states is there larger than at $n=5$.\cite{notespin}

\section{ Magnetic phase diagrams }

The instability of the system towards either F or AF order (\ref{chi}) is 
driven by the value of the Stoner parameter $I_d$ ($I_s$). We illustrate 
this by considering the phases with a uniform and with a two-sublattice 
magnetic structure,
\begin{equation}
\langle n_{i\alpha,\sigma}\rangle=
\frac{1}{2}\left[ n_0 + \lambda_{\sigma}m   
 + ( \eta + \lambda_{\sigma}\nu ) e^{i{\bf Q}{\bf R}_i} \right],
\label{density}
\end{equation}
where $m$, $\eta$ and $\nu$ are order parameters, and $\lambda_{\sigma}
=\pm 1$ for $\sigma=\uparrow,\downarrow$. The quasiparticles 
are given (up to constant energy shifts) by 
\begin{equation}
E^{\pm}_{{\bf k}\alpha,\sigma}=
\frac{1}{2}(\varepsilon_{{\bf k}+{\bf Q}}+\varepsilon_{\bf k})
\pm \frac{1}{2}\left[(\varepsilon_{{\bf k}+{\bf Q}}-\varepsilon_{\bf k})^2
+\Delta_{\sigma}^2\right]^{1/2}, 
\label{quasi}
\end{equation}
and ${\bf Q}=(\pi,\pi)$ is the nesting vector at $R=0$. The value of the gap 
at half-filling ($n_0=1$) is given by the effective interaction, 
$\Delta_{\sigma}=\Delta=I_d\nu/2$ in the $d$-band case 
($\Delta_{\sigma}=\Delta=I_s\nu/2$ in the $s$-band case). 
 
The alternating magnetic order results in a two-sublattice magnetic structure
and opens a gap $\Delta_{\sigma}$ for $\sigma$-spin electrons. Assuming the
filling by $n_0$ electrons (per one band $d$-subband or $s$-band), 
we analyze only the following commensurate magnetic phases: 
  (i) ferromagnetic           (F), $m=\min\{n_0,2-n_0\}$, $\nu=\eta=0$; 
 (ii) partial ferromagnetic  (PF), $m\not=0$, $m<\min\{n_0,2-n_0\}$, 
                                   $\nu=\eta=0$; 
(iii) antiferromagnetic      (AF), $\nu\not=0$, $m=\eta=0 $; 
 (iv) special ferrimagnetic (SFI), $m=|1-n_0|$, $\nu\not=0$, $\eta\not=0$;
      here the stability follows from the Fermi level lying within the gap 
      between two majority Slater subbands. 

In the region of their stability, the energies of magnetic phases are 
determined using the total energy expressions within the GHF,
\begin{equation}
E(\{n_{i\alpha,\sigma}\})=E_{\rm GHF}+E_{corr},
\label{etotal}
\end{equation}
where $E_{\rm GHF}$ is determined as in HF approximation from the 
quasiparticle energies (\ref{quasi}). For simplicity, we give only the 
formula for less than half-filling ($n_0\leq 1$),  
\begin{equation}
E_{\rm GHF}={1\over N}\sum_{\alpha,{\bf k}\in K(  \uparrow)}
                       E^{-}_{{\bf k}\alpha,  \uparrow}
       +{1\over N}\sum_{\alpha,{\bf k}\in K(\downarrow)}
                       E^{-}_{{\bf k}\alpha,\downarrow}
       -\langle H_{int}\rangle,
\label{eghf}
\end{equation}
where $K(\sigma)$ is a set of the occupied quasiparticle states 
$E^-_{{\bf k}\alpha,\sigma}$ (\ref{quasi}) in the lower Slater subband for 
$\sigma$-spin. The interaction energy $\langle H_{int}\rangle$ is subtracted 
to avoid the double counting, with the form of $H_{int}$ determined by the 
used Hamiltonian, either $H_s$ or $H_d$. 

The correlation energy depends on the magnetic order, 
and is calculated using the local ansatz \ref{ansatz},\cite{Ole83,Sto96}
\begin{equation}
E_{corr}=
{\langle \Psi_0(\{n_{i\alpha,\sigma}\})|H_d|
         \Psi_0(\{n_{i\alpha,\sigma}\})\rangle
\over \langle \Psi_0(\{n_{i\alpha,\sigma}\})| 
              \Psi_0(\{n_{i\alpha,\sigma}\})\rangle}
-E_{\rm GHF},
\label{ecorr}
\end{equation}
In the $s$-band case we adopted the value of $E_{corr}=0$, which is exact 
for the F states, and avoids double counting of the correlation contributions 
in the P and AF states. The treatment of atomic correlations in the $d$-band
beyond the GRPA is only approximate, but suffices to get qualitative 
results for the magnetic phase diagrams reported below. These correlations 
vanish in the F phase, but are finite in the P state, and therefore the value 
of the Stoner parameter at the F instability is found in an approximation. On 
the contrary, they have no influence on the lines of the AF instabilities, 
where the order parameter $\nu$ increases continuosly from zero. Here we used 
the local approximation to evaluate the respective averages, when the 
exponentials in the wave functions $\Psi_0(\{n_{i\alpha,\sigma}\})$ 
(\ref{ansatz}) are expanded in Eq. (\ref{ecorr}). More details may be found 
in Refs. \onlinecite{Ole83}. The magnetic phase diagrams are next found using 
the instabilities of the nonmagnetic states, and by comparing the total 
energies (\ref{etotal}) of the magnetic phases in the region of their 
stability.   

\begin{figure}
\centerline{\psfig{figure=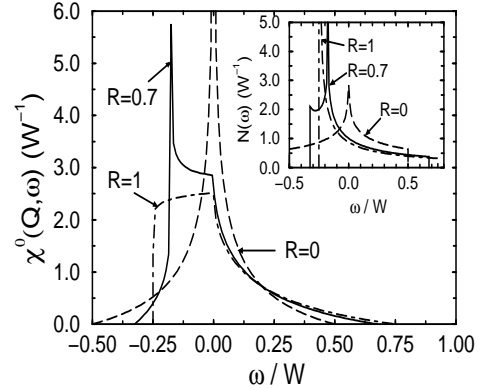,height=2.0in,width=2.4in}}
\narrowtext
\smallskip
\caption
{Free susceptibility $\chi^0({\bf Q},\omega)$ as obtained at the nesting
vector ${\bf Q}=(\pi,\pi)$ for the 2D densities of states with $R=0$, $0.7$,
and $1.0$ ($W=8t$). The corresponding densities of states $N(\omega)$ are 
shown in the inset.}
\label{chi+dos}
\end{figure}
\begin{figure}
\centerline{\psfig{figure=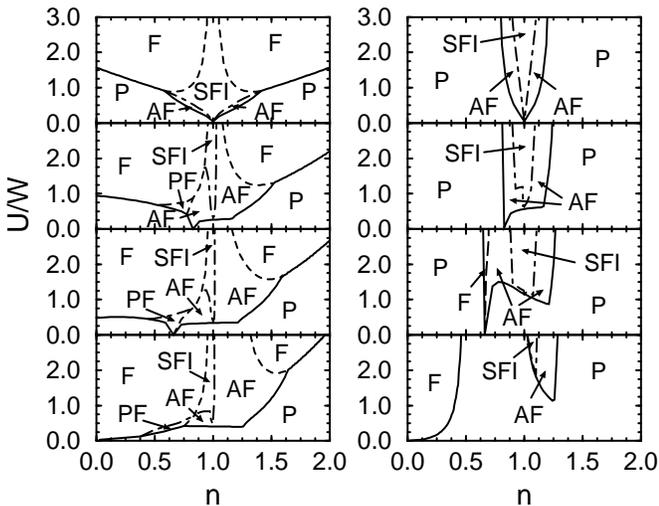,height=2.6in,width=3.4in}}
\narrowtext
\smallskip
\caption
{RPA (left) and GRPA (right) phase diagrams for the $s$-band case and 
for $R=0$, 0.4, 0.7, and 1.0 from top to bottom.
Full lines give the instabilities with increasing 
$U$ of paramagnetic (P) to ferromagnetic (F) or antiferromagnetic (AF) 
states. The special ferrimagnetic states (SFI) are separated by dot-dashed 
lines. Ferromagnetic states are saturated (F) except for small regions in 
the three lowest panels to the left (PF).}
\label{sband}
\end{figure}

As an illustrative example, we limit ourselves here to the 2D models 
(\ref{shubbard}) and (\ref{dhubbard}) with nearest ($t$) and next-nearest 
($t'$) neighbor hopping. By changing the value of $R=2t'/t$, the bands 
(\ref{dispersion}) become flat and the noninteracting susceptibilities 
$\chi^0({\bf k},\omega)$ change. The singularity in the AF susceptibility 
$\chi^0({\bf Q},\omega)$ [${\bf Q}=(\pi,\pi)$] moves to lower energies, as 
shown in Fig. \ref{chi+dos}, and gradually disappears, while the FM 
susceptibility, $\chi^0({\bf 0},\omega)=N(\omega)$, develops a peak at low 
energies (see the inset of Fig. \ref{chi+dos}). Therefore, the RPA 
instabilities of the $s$-band model towards FM states occur at all electron 
densities (Fig. \ref{sband}), and are enhanced at low $n$ for increasing 
$t'$.\cite{Lin87} One finds that these instabilities occur towards the 
saturated FM states (with a maximum value of $m=|1-n_0|$) in most cases. As 
expected, the AF (SFI) order is more stable near half-filling. 

If correlation effects (renormalization of $U$ within the GRPA) 
are included, the phase diagrams change drastically (Fig. \ref{sband}). 
The FM instability disappears almost entirely in the GRPA if $R<1$, except 
just in a narrow region around the VHS for larger values of $R$. At $R=0$ 
it is suppressed at any $n$, in agreement with Rudin and 
Mattis.\cite{Rud85} As also no instability of the P state towards the FM 
order is found in infinite dimension,\cite{Fre95} it is likely that it does 
not exist in hypercubic lattices at $R=0$.\cite{notefm} The situation changes 
when particle-hole symmetry is broken for $t'>0$, and the FM order is 
stabilized by infinitesimal $U$ at the VHS point, if $R>0.55$.\cite{Hlu96} 
Remarkably, at $R=1$ {\it ferromagnetism is stable at low density}, $n<0.45$, 
as a consequence of flat band behavior, $\varepsilon_{\bf k}=-2t$, if 
${\bf k}=(k_x,0)$ or ${\bf k}=(0,k_y)$. 

The AF order is found to be more robust and exists around half-filling at 
any value of $R\leq 1$, with electron correlations included. In agreement 
with expectations, the region of AF order at intermediate values of 
$U/W\geq 1.0$ expands when the kinetic energy in the ordered state increases 
with the increasing intrasublattice hopping $t'$, both in RPA and in GRPA,
as shown in Fig. \ref{sband}. On the contrary, at $R=1$ the region of 
antiferromagnetism is much reduced compared with $R=0$. It is expected that 
incommensurate magnetic order with AF correlations is stable in between 
the AF and P phases, as shown recently within this method in infinite 
dimension.\cite{Fre95,Tah97} 

\begin{figure}
\centerline{\psfig{figure=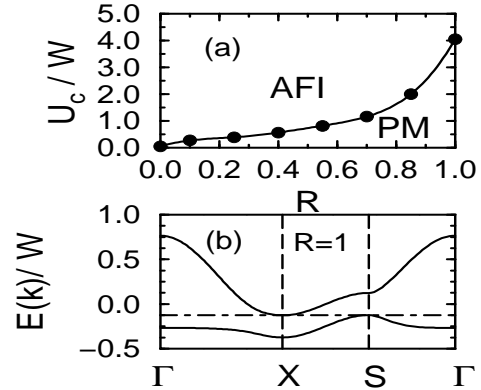,height=2.0in,width=2.4in}}
\narrowtext
\smallskip
\caption
{Instability of a paramagnetic metallic (PM) to antiferromagnetic 
insulating (AFI) ground state at half-filling ($n=1$) in the Hubbard model
(1): (a) critical value of the Coulomb interaction $U_c$ as function of 
$R=2t'/t$; (b) electronic structure in an almost insulating
(weakly metallic) AF state obtained at $U=4.04W$ and $R=1$.}
\label{afi}
\end{figure}

In contrast to ferromagnetism, the tendency towards AF order at and near 
half-filling is weakened by the increasing next-nearest neighbor hopping 
$t'$, as the shape of the Fermi surface changes and the perfect nesting 
condition is not satisfied. At $R=0$, the transition to the AF state occurs 
at infinitesimal $U>0$, and the magnetic moment $\nu$ increases gradually 
from zero with increasing $U$. This behavior is characteristic of perfect 
nested band structures, and is replaced by a jump to a finite magnetization 
$\nu$ which occurs at finite $U>0$, if $R>0$. With increasing $R$ the lower 
quasiparticle subband $E^-({\bf k})$ (\ref{quasi}) becomes more flat and it 
is gradually more difficult to stabilize an AF state. Therefore, the critical 
value of the Coulomb interaction $U_c/W$, increases with increasing $R$, as 
shown in Fig. \ref{afi}. 

It is interesting to ask whether the insulating AF state at half filling and
$R=0$ is replaced by itinerant antiferromagnetism with increasing values of
$t'$. The gap in the AF state at the transition point is determined by the 
energy difference between the top of the lower Slater subband $E^-$ at 
${\bf k}=(\pi/2,\pi/2)$, and the minimum of the upper Slater subband $E^+$ 
at ${\bf k}=(\pi,0)$. Using the effective Coulomb interaction ${\bar U}$, we 
find that the AF state is always insulating at the transition point, except 
at $R=1$, where the gap between the Slater subbands closes and the AF state 
is weakly metallic, as shown in Fig. \ref{afi}. The onset of metallic 
behavior occurs in GRPA at $2t'=\Delta$, where $\Delta_{\sigma}=\Delta=
I_s\nu/2$ in Eq. (\ref{quasi}). If $R=1$, one finds at the transition point 
${\bar U}_c=0.409W$, and $\nu=0.606$, and the splitting $\Delta=0.124W$, just 
somewhat smaller than $2t'=0.125W$. This results in a very small difference 
of the energies, $\Delta E=E^+(X)-E^-(S)=-2.2\times 10^{-3}W$, and supports 
the recent conclusion of Duffy and Moreo\cite{Duf97} that itinerant 
magnetism is difficult to realize in the $U-t-t'$ Hubbard model, without the 
second neighbor hopping $t^{''}$ along the $x$ and $y$ directions. 
Morever, it seems that the critical values of $U$ obtained for the onset of 
AF LRO might be overestimated at finite $R$, as we find the AF instability 
at $U_c\simeq 4.5t$ for $R=0.4$, while the quantum Monte-Carlo calculations 
suggest a value $U_c\simeq 2.5t$.\cite{Lin87,Duf97} However, they agree
with the metal-insulator transition estimated to be somewhere in the range 
of $4t<U_c<6t$, if $R=0.4$.

\begin{figure}
\centerline{\psfig{figure=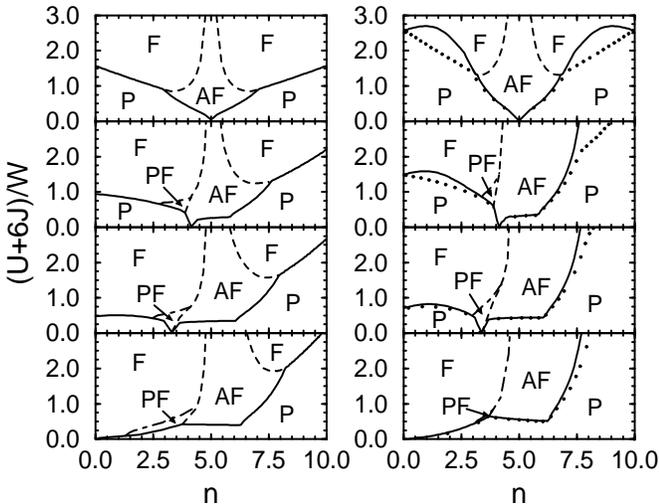,height=2.6in,width=3.4in}}
\narrowtext
\smallskip
\caption
{The same as in Fig. \protect{\ref{sband}}, but for the $d$-band model (2) 
with $J/U=0.25$. The magnetic instabilities are shown bu full lines, while
dotted lines indicate the same instabilities if the spin-spin correlations 
are neglected.}
\label{dband}
\end{figure}

The RPA phase diagram of the $d$-band model is similar to that of the 
$s$-band (Fig. \ref{dband}), but the SFI phase is destabilized by the 
interorbital exchange interaction $J$, and thus the AF state is found 
instead in a broader range of parameters. As a consequence of the weaker 
screening by particle-particle diagrams (\ref{urn}), the conditions for FM 
LRO are less restrictive,\cite{notekk} and the FM instabilities occur in the 
GRPA at any value of $R$, but for larger interactions (Fig. \ref{dband}). 
As in the $s$-band case, FM order is favored at low electronic filling for 
$0.4<R<1.0$ due to weaker dispersion. As already seen in the screening of 
the Stoner parameter $I_d$, the spin-spin correlations restrict the region 
of FM states, in particular at $R=0$, and for $n>7.5$, if $R\ge 0.4$. 
Furthermore, one finds that weak ferromagnetism is somewhat more pronounced 
than in the $s$-band and survives for the screened interactions, but occurs 
only in a relatively narrow range of $n$. This shows that partly polarized 
FM states are more likely to result from either the local maxima or the 
splittings between $e_g$ and $t_{2g}$ orbitals in realistic band structures 
of $3d$ transition metals. 

\section{ Summary and conclusions }

Summarizing, we conclude that the FM states are stable not only at, but
also in the neighborhood of the VHS in the $s$-band model, if the band 
filling $n$ is small. This suggests that ferromagnetism might be realized
in the Hubbard model also in three dimensions, but only if the particle-hole 
symmetry is broken and the kinetic energy satisfies rather {\em extreme} 
conditions. Therefore, the $s$-band model cannot serve as a generic 
description of itinerant magnetism in transition metals. Instead the orbital 
degeneracy of the $d$-states is crucial to explain magnetic states, and the 
interorbital exchange coupling $J$ plays an important role.

Although no more than qualitative statements can be made for realistic
transition metals, it is interesting to compare the obtained phase diagrams
with the known interaction parameters of $3d$ systems. The values of $U$ and
$J$ are approximately known for $3d$ transition metals and may be obtained
from the multiplet splittings, as discussed in detail by van der Marel and
Sawatzky.\cite{vdM88} This analysis leads to $U=2.55$, $2.76$, and $2.97$ eV,
for Fe, Co, and Ni, respectively, 
while the values of $J$ are given by the same ratio, $J/U=0.27$.\cite{Sto90}
These values of $U$ are close to those considered by one of us in an earlier 
study of ferromagnetism in $3d$ metals,\cite{Sto90} and we treat them here as 
representative ones. Taking the bandwidths of Fe, Co, and Ni as determined
by Andersen, Jepsen, and Gl\"otzel:\cite{And85} $W=5.43$, $4.84$, and $4.35$
eV, respectively, this results in the HF values of the magnetic interaction
for these elements of $(U+6J)/W=1.23$, $1.49$, and $1.79$, respectively.
We assumed the electron filling of $n=7.3$, $8.3$, and $9.4$ in the $d$-band
and $R=0$ in our model (\ref{dhubbard}) to simulate qualitatively the
situation in Fe, Co, and Ni, and found a reduction of the respective Stoner 
parameter due to particle-particle renormalization and the spin-spin 
correlations of the order of $I_d/I_{\rm HF}\simeq 0.61$, $0.58$, and $0.54$, 
respectively. Interestingly, these values are almost constant as the 
increasing values of $(U+6J)/W$ are counterbalanced by the smaller reduction 
of $I_d$ (both by particle-particle and even more by spin-spin correlations) 
when the electronic filling increases towards the filled band. Taking 
$J/U=0.25$ in a realistic model with canonical $d$ bands,\cite{Sto96} 
one finds instead that $I_d/I_{\rm HF}\simeq 0.63$, $0.57$, and $0.52$. 
The agreement is very good indeed granted the simplicity of the model. 

Furthermore, using the values of $I_d$ derived above and realistic values of 
$d$-bandwidths,\cite{And85} one finds the Stoner parameters: $I_d\simeq 
0.72$, $0.82$, and $0.92$ eV for Fe, Co, and Ni, respectively. These values 
are significantly lower than those deduced by Gunnarsson from the local 
density approximation (LDA) calculations, being $0.92$, $0.99$, and $1.01$, 
respectively.\cite{Gun76} In spite of the qualitative nature of this 
comparison, this allows us to conclude that important corrections of the 
Stoner parameter exist due to nonlocal electron correlation effects, as in
particular due to the spin-spin interorbital correlations, which cannot be 
dealt with in the standard band structure calculations based on the LDA. 
Thus, in spite of some attempts which exist in the literature,\cite{Sev93} 
there is no way to deduce reliable values of the Stoner parameter from band 
structure calculations performed within the LDA. 

An additional factor which might stabilize ferromagnetism in the $d$-band 
model is the flattening of the bands with increasing values of $R$. Then 
magnetic states are possible even for rather small interactions $U$. Of 
course, this tendency is overemphasized in the present 2D model by the Van
Hove singularity, but it is expected that strong next-nearest neighbor 
hopping $t'$ may lead to FM instabilities also in three-dimensional 
or quasi 2D systems at low filling. It would be interesting to verify this
prediction, if such materials could be synthesized.

Altogether, we have shown that a correct quantitative description of 
ferromagnetism in transition metals is only possible within a realistic 
$d$-band model (\ref{dhubbard}), and when the particle-particle screening 
and spin-spin correlations are included. In order to obtain more 
quantitative results, however, realistic densities of states have to be 
used. A somewhat different situation, however, is found for the AF states, 
where the band structure effects (like perfect nesting) dominate, at least 
at weak and intermediate values of $U$, and the $s$-band model might then 
suffice to explain qualitatively the experimental data for AF Cr and its 
alloys.\cite{Fre95}

\section*{ Acknowledgments }

We thank Richard Hlubina and Gernot Stollhoff for stimulating discussions. 
A.M.O. acknowledges the partial support by the Committee of Scientific 
Research (KBN) of Poland, Project No.~2 P03B 144 08.

\end{multicols} 


\begin{thebibliography}{99}
\bibitem{noteid}
                The intraorbital (particle-particle) correlations were 
                described by the local ansatz (\protect{\ref{ansatz}}) 
                in Ref. \protect{\onlinecite{Sto90}}.
\bibitem{notespin}
                The correlation energy gained by spin correlations is 
                largest at half-filling and vanishes if $n\to 0$.
\bibitem{notefm}
                The absence of F instability in the presence of correlation 
                effects does not suffice to conclude that ferromagnetism 
                does not occur at any density and $U$. As shown in Refs. 
                \protect{\onlinecite{vdL91,Han93}}, the F GS is locally
                stable in a 2D Hubbard model near half-filling at 
                $U\to\infty$.
\end{thebibliography}
\end{document}